\documentclass[preprint2]{aastex}

\begin{document}
\title{OH 18 cm Transition as a Thermometer for Molecular Clouds}
\author{\
Yuji Ebisawa\altaffilmark{1},
Hiroshi Inokuma\altaffilmark{1},
Nami Sakai\altaffilmark{2},
Karl M. Menten\altaffilmark{3},
Hiroyuki Maezawa\altaffilmark{4}
\\and
Satoshi Yamamoto\altaffilmark{1}}

\altaffiltext{1}{Department of Physics, The University of Tokyo, Hongo, Bunkyo-ku, Tokyo 113-0033, Japan}
\altaffiltext{2}{RIKEN, 2-1 Hirosawa, Wako, Saitama 351-0198, Japan}
\altaffiltext{3}{Max-Planck-Institut f\"{u}r Radioastronome, Auf dem H\"{u}gel 69, D-53121 Bonn, Germany}
\altaffiltext{4}{Department of Physical Science, Osaka Prefecture University, 1-1 Gakuen-cho, Naka-ku, Sakai, Osaka 599-8531}
\begin{abstract} 
We have observed the four hyperfine components of the 18 cm OH transition toward the translucent cloud eastward of Heiles Cloud 2 (HCL2E),
the cold dark cloud L134N, and the photodissociation region of the $\rho$-Ophiuchi molecular cloud with the Effelsberg 100 m telescope.
We have found intensity anomalies amongst the hyperfine components in all three regions.
In particular, an absorption feature of the 1612 MHz satellite line against the cosmic microwave background has been detected toward HCL2E and two positions of the $\rho$-Ophiuchi molecular cloud.
On the basis of statistical equilibrium calculations, we find that the hyperfine anomalies originate from the non-LTE population of the hyperfine levels,
and can be used to determine the kinetic temperature of the gas over a wide range of H$_2$ density (10$^2$ -- 10$^7$ cm$^{-3}$).
Toward the center of HCL2E, the gas kinetic temperature is determined to be 53$\pm$1 K, and it increases toward the cloud peripheries ($\sim$ 60 K).
The ortho-to-para ratio of H$_2$ is determined to be 3.5 $\pm$ 0.9 from the averaged spectrum for the 8 positions.
In L134N, a similar increase of the temperature is also seen toward the periphery.
In the $\rho$-Ophiuchi molecular cloud, the gas kinetic temperature decreases as a function of the distance from the exciting star HD147889.
These results demonstrate a new aspect of the OH 18 cm line as a good thermometer of molecular cloud envelopes.
The OH 18 cm line can be used to trace a  new class of warm molecular gas surrounding a molecular cloud, which is not well traced by emission of CO and its isotopologues.
\end{abstract}
\section{INTRODUCTION} 
\label{introduction}

\begin{figure}
  \epsscale{0.5}
  \plotone{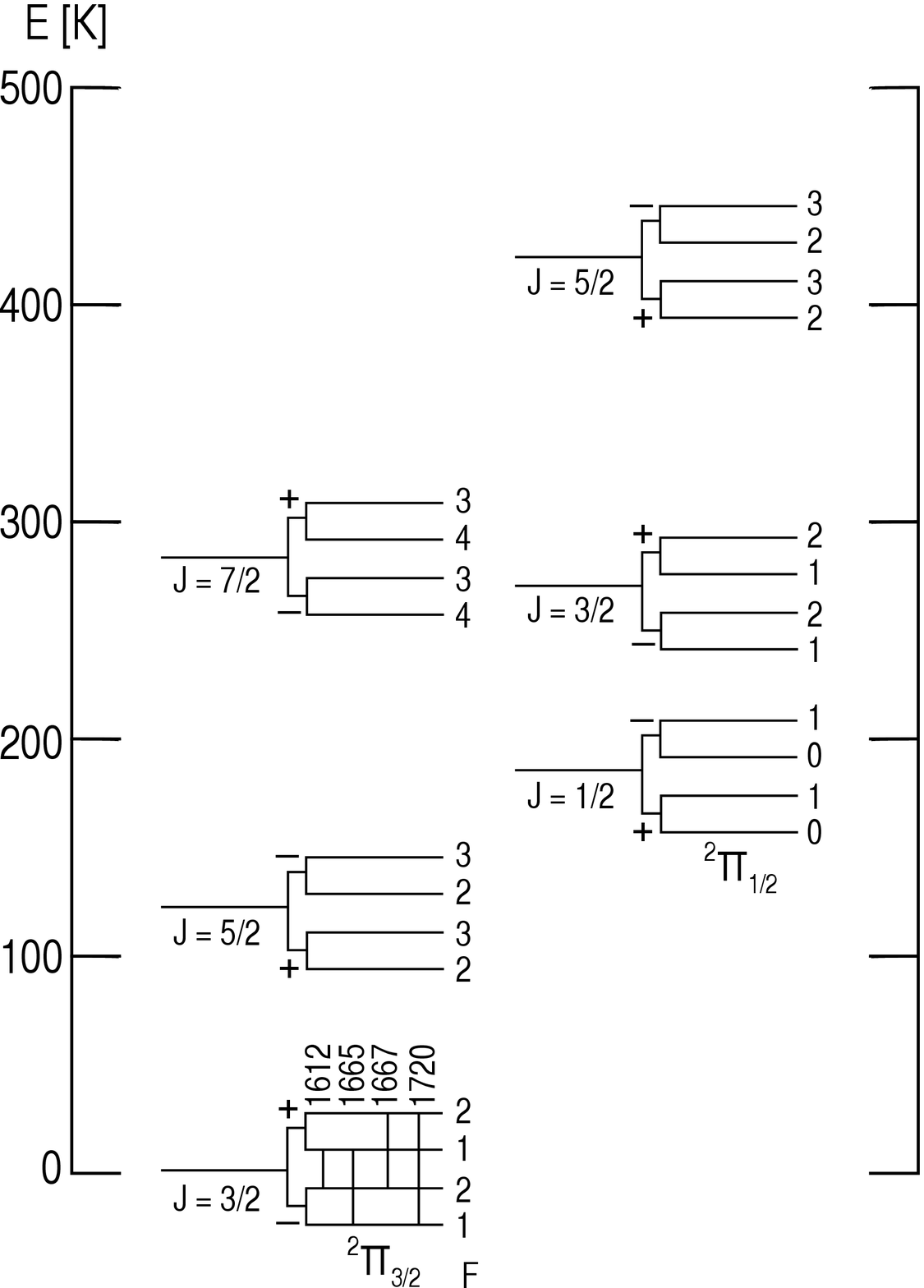}
  \caption{The rotational energy level structure of OH.
    The fine and hyperfine structure levels are schematic.
  \label{fig_OH_energy}}
\end{figure}

The 18 cm transition of the hydroxyl radical (OH) was the first spectral line from interstellar molecules detected in the radio wavelength region (Weinreb et al. 1963).
It has extensively been observed toward diffuse clouds, molecular clouds, star-forming regions, supernova remnants, late-type stars, and external galaxies.
It is the fine structure transition between the $\Lambda$-type doubling levels in the lowest rotational state ($J$=3/2,$\Omega$=3/2), and consists of four hyperfine components at frequencies near 1612 MHz, 1665 MHz, 1667 MHz, and 1720 MHz due to the nuclear spin of the hydrogen nucleus (Figure $\ref{fig_OH_energy}$, Table $\ref{table_OH}$).
The 1665 MHz and 1667 MHz lines are called main lines, whereas the 1612 MHz and 1720 MHz lines are called satellite lines.
It is well known that the relative populations of the hyperfine structure (hfs) levels can attain values that deviate
from intrinsic values attained in Local Thermodynamic Equilibrium (LTE), i.e., 1612:1665:1667:1720 MHz = 1:5:9:1.
This is caused by an interplay of excitation and de-excitation through rotationally excited states ($J$=5/2,$\Omega$=3/2 and $J$=1/2,$\Omega$=1/2) (e.g.\ Elitzur 1976; van Langevelde et al. 1995).
In fact, the 18 cm transition of OH often displays maser emission in star-forming regions (mainly the 1665/1667 lines),
supernova remnants (only the 1720 MHz line),
and the circumstellar envelopes of mass losing evolved stars (mainly the 1612 MHz line)
(e.g. Gundermann et al. 1965; Weaver et al. 1965; Cohen 1995). 
Furthermore, so-called $``$conjugate$"$ behavior of the 1612 MHz and 1720 MHz satellite lines, which appear in absorption and emission, respectively, has been reported for sources with bright background continuum
(see, e.g.\ van Langevelde et al. 1995; Weisberg et al. 2005; Kanekar et al. 2004).
\par
Observations of the OH 18 cm transition toward quiescent molecular clouds have also been conducted extensively (e.g.\ Heiles 1968; Cudaback \& Heiles 1969; Crutcher 1973; Myers et al. 1978).
Mattila et al. (1979) carried out observations including the satellite lines toward L134, and reported the 1612 MHz line to be fainter than the 1720 MHz line.
Even more extreme differences between the latter two lines were observed toward the HII region W40 by Crutcher (1977) who found the 1612 MHz line shows absorption at velocities at which the 1720 MHz line shows emission.
This $``$conjugation$"$ of these two lines was also observed toward the diffuse gas in the line of sight toward the active galactic nucleus Centaurus A (van Langevelde et al. 1995).
In addition, the Zeeman effect of the OH 18 cm transition has been conducted to measure the magnetic field strength of molecular clouds (e.g. Crutcher et al. 1993; Troland \& Crutcher 2008).
\par
As for astrochemical interest, Turner \& Heiles (1974) investigated the relation of the OH abundance to the visual extinction in two regions of the Taurus molecular cloud.
Harju et al. (2000) conducted sensitive mapping observations of the OH 18 cm transition toward the filamentary Taurus Molecular Cloud-1 (TMC-1) in HCL2 with the Effelsberg 100 m telescope.
This filament is famous for showing significant chemical differentiation.
Carbon-chain molecules are abundant in the south-eastern part of the filament, whereas NH$_3$ is abundant in the north-western part (e.g., Little et al. 1979; Hirahara et al. 1992).
This chemical differentiation is naturally explained in terms of an chemical evolutionary effect (Hirahara et al. 1992; Suzuki et al. 1992).
Harju et al. (2000) investigated the OH/C$^{18}$O abundance ratio along the ridge from such a chemical point of view, and found that it is almost constant over the ridge.
This indicates almost constant abundance of OH during cloud evolution.
At the same time, they reported the intensity anomaly of the hyperfine components.
The 1612 MHz and the 1720 MHz lines, both in emission, are slightly brighter and fainter, respectively, than expected for LTE.
They argued that this anomaly is due to the contribution of the far IR continuum emission from dust.
However, the observed region is limited to the small area of the TMC-1 ridge, and more observations are required to investigate the origin of the anomaly.  
\par
In the course of our OH 18 cm transition survey toward the whole HCL2 region with astrochemical interest, we fortuitously found the conjugate behavior,
where the 1612 MHz line is observed in absorption and the 1720 MHz line shows enhanced emission.
This anomaly is different from that found by Harju et al. (2000), and is similar to that found toward L134 and other molecular clouds mentioned above.
Although such anomalies have been discussed in terms of non-LTE effect (Elitzur 1976; Mattila et al. 1979),
detailed analyses have not been reported because of a lack of high-quality observational data of the satellite lines.
In this paper, we demonstrate that this hyperfine anomaly can be used as a good thermometer of molecular clouds over a relatively wide range of the H$_2$ density.
The whole survey of OH in the HCL2 cloud will be published separately.

\section{OBSERVATIONS}
\begin{figure}
  \epsscale{0.8}
  \plotone{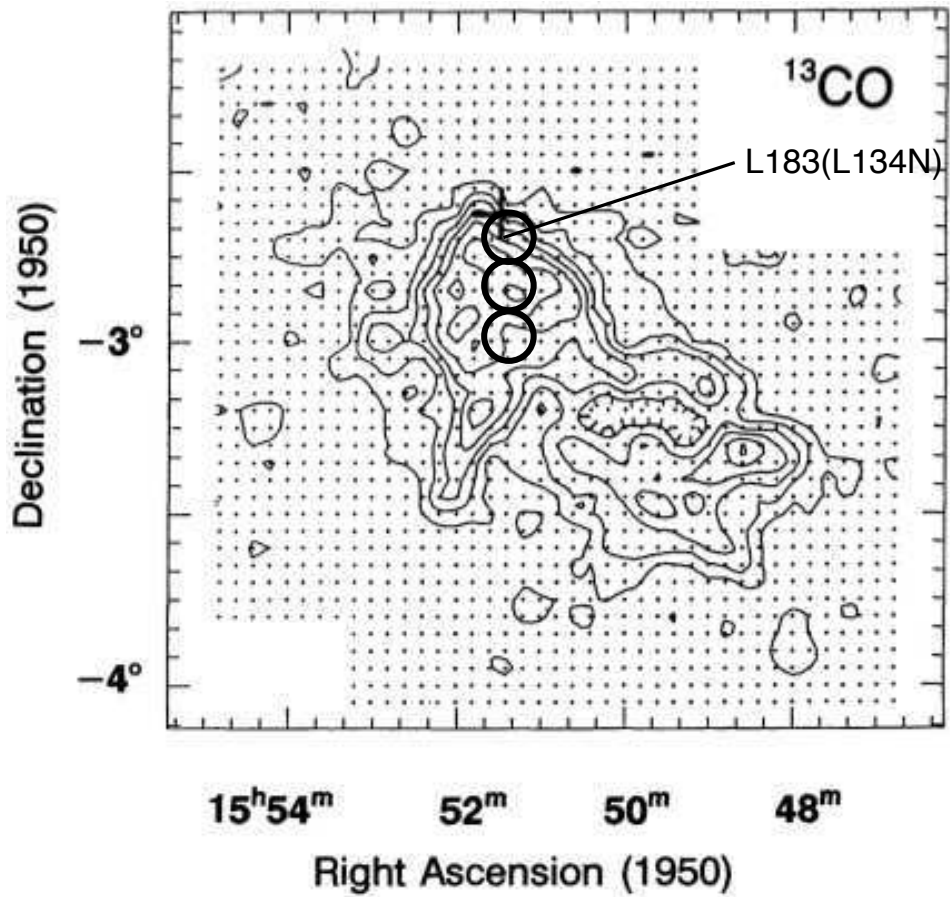}
  \caption{Contours show the integrated intensity map of $^{13}$CO ($J$=1-0) toward L134N region reported by Laureijs et al. (1995). Circles represent the positions observed in OH.
    The cross mark represents the position of the NH$_3$ core of L134N (L183) indicated by Laureijs et al. (1995).
    \label{fig_map_L183}}
\end{figure}

\begin{figure}
  \epsscale{0.8}
  \plotone{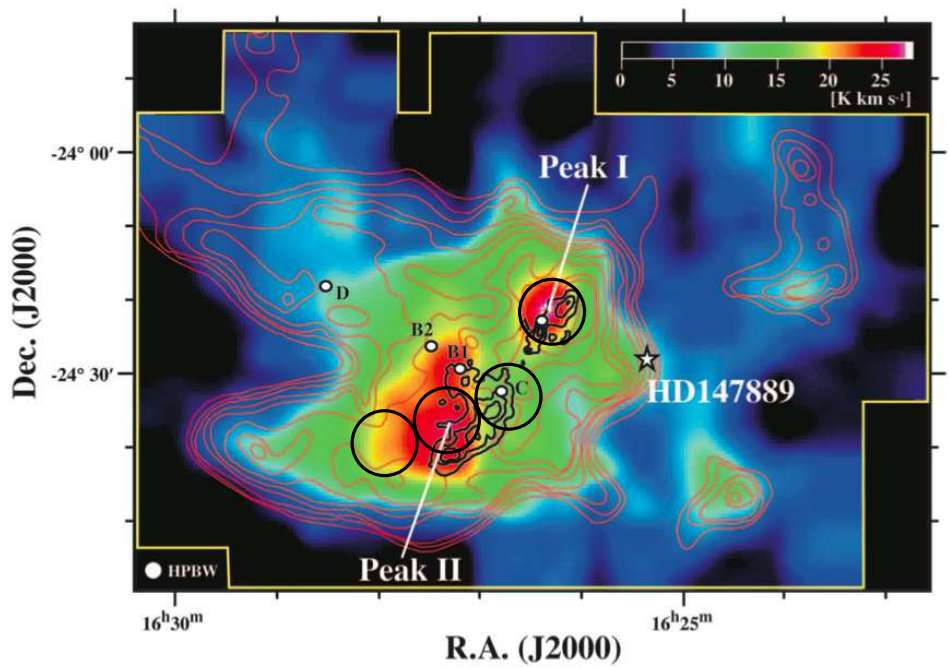}
  \caption{The $\rho$-Oph region: Black and red contours shows the integrated intensity map of C$^{18}$O ($J$=1-0) and {}$^{13}$CO ($J$=1-0) lines respectively, overlaid on color image of the integrated intensity map of the [C I] (${}^3P_1-{}^3P_0$) emission (Kamegai et al. 2003). Large black circles represent the positions observed in OH.
    The position of the heating and ionizing star HD147889 is indicated by a star.
    \label{fig_map_Oph}}
\end{figure}

Observations were carried out with the MPIfR 100 m telescope at Effelsberg in 2009 September and 2013 October.
We observed four hyperfine component lines of the ground-state $\Lambda$-type doubling transition of OH (Table $\ref{table_OH}$).
The FWHM beam width of the telescope is $8.2'$.
We used the 18 cm/21 cm prime focus receiver as a frontend, whose system noise temperature is about 20 K.
The telescope pointing was checked every three hours by observing nearby continuum sources, and was maintained to be better than 20$''$.
The observations were carried out with the frequency-switching mode, where the frequency offset is 0.1 MHz.
We used Fast Fourier Transform spectrometer (FFTS) as a backend, whose bandwidth and resolution are 100 MHz and 3.5 kHz
(corresponding to 0.56 kms$^{-1}$ at 1667 MHz), respectively.
In this paper, we report the results obtained toward the translucent cloud at the eastern side of HCL2 (Figure $\ref{fig_map_HCL2E}$),
the cold dark cloud L134N (Figure $\ref{fig_map_L183}$),
and the $\rho$-Ophiuchi molecular cloud (Figure $\ref{fig_map_Oph}$).

\section{RESULTS}
\subsection{HCL2E}
\begin{figure}
  \epsscale{0.8}
  \plotone{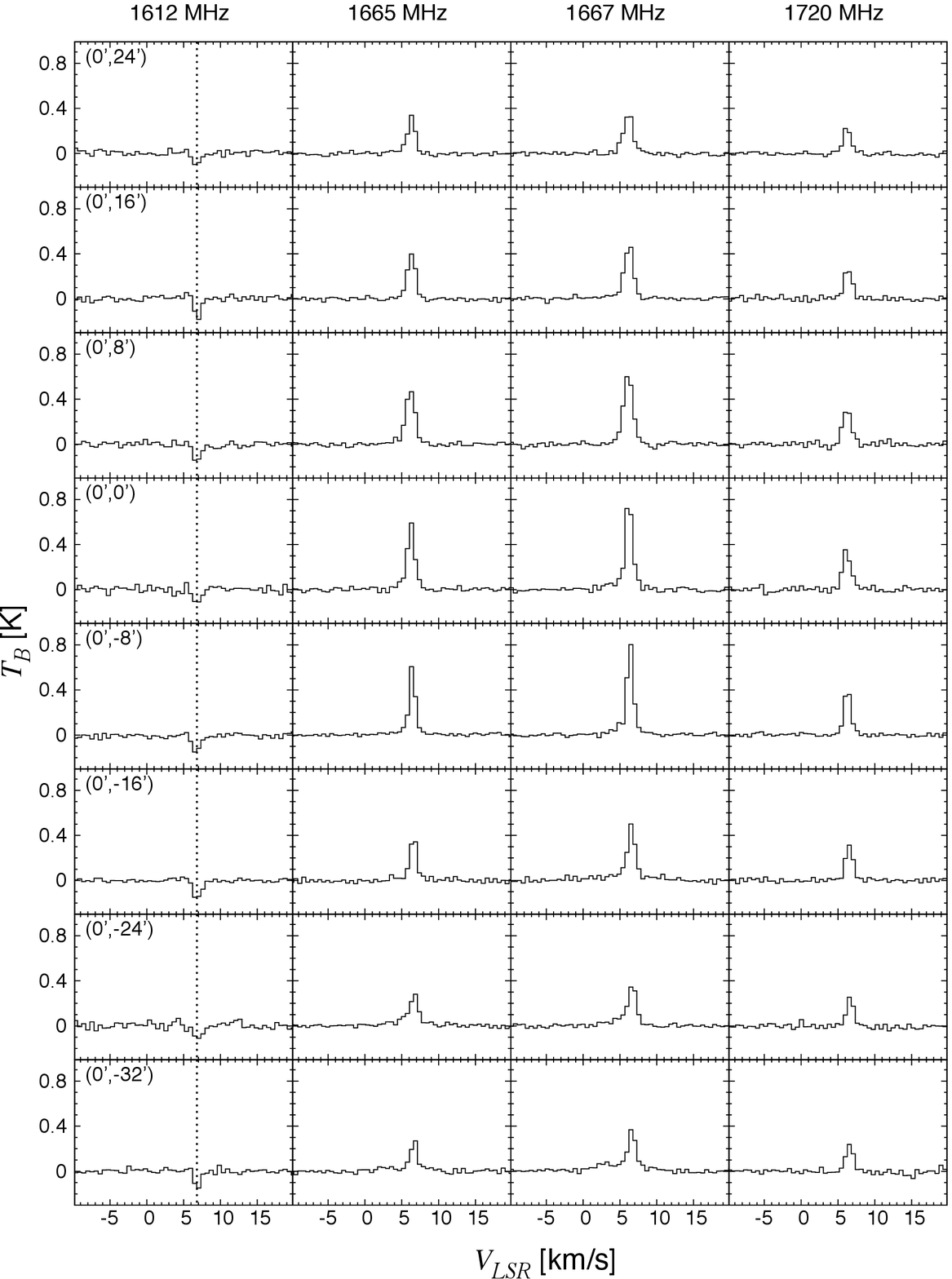}
  \caption{Observed spectra of the OH 18 cm transition toward HCL2E. A vertical dashed line represents the LSR velocity of the 1612 MHz line.
    \label{fig_spectrum_HCL2E}}
\end{figure}

We first found the absorption feature in the 1612 MHz line toward a translucent cloud located at 1.5$^{\circ}$ east of HCL2
(referred hereafter as HCL2E) (Figure $\ref{fig_map_HCL2E}$) during the 2009 observing run.
HCL2E has an apparent size of about $30'$, and is connected to HCL2 in {}$^{13}$CO ($J$=1-0) emission (Mizuno et al. 1995; Narayanan et al. 2008) and [C I] emission (Maezawa 2000).
A relatively high C/CO ratio further indicates the translucent nature of this cloud,
where the visual extinction is estimated to be about 4 visual magnitudes (Maezawa 2000).
We observed the four hyperfine components of the OH 18 cm transition toward 8 positions with $8'$ spacing along the strip line from north to south indicated in Figure $\ref{fig_map_HCL2E}$.
The result is shown in Figure $\ref{fig_spectrum_HCL2E}$.
The absorption feature in the 1612 MHz transition is evident for all the positions.
Since no radio continuum source is known toward HCL2E, this feature represents absorption against the cosmic microwave background.
Furthermore, the 1720 MHz line is observed in emission with an intensity much stronger than expected from LTE.
\par
The absorption in the 1612 MHz component and enhanced emission from the 1720 MHz component have been reported for a few other sources as mentioned in Section 1
(e.g.\ Crutcher 1977; van Langevelde et al. 1995; Weisberg et al. 2005; Kanekar et al. 2004).
It is known that this occurs by collisional excitation to rotationally excited states and subsequent radiative relaxation to the ground rotational state.
Our observations further confirm that such conjugation occurs in quiescent clouds without bright background continuum emission.

\subsection{L134N} 
\begin{figure}
  \epsscale{1.1}
  \plotone{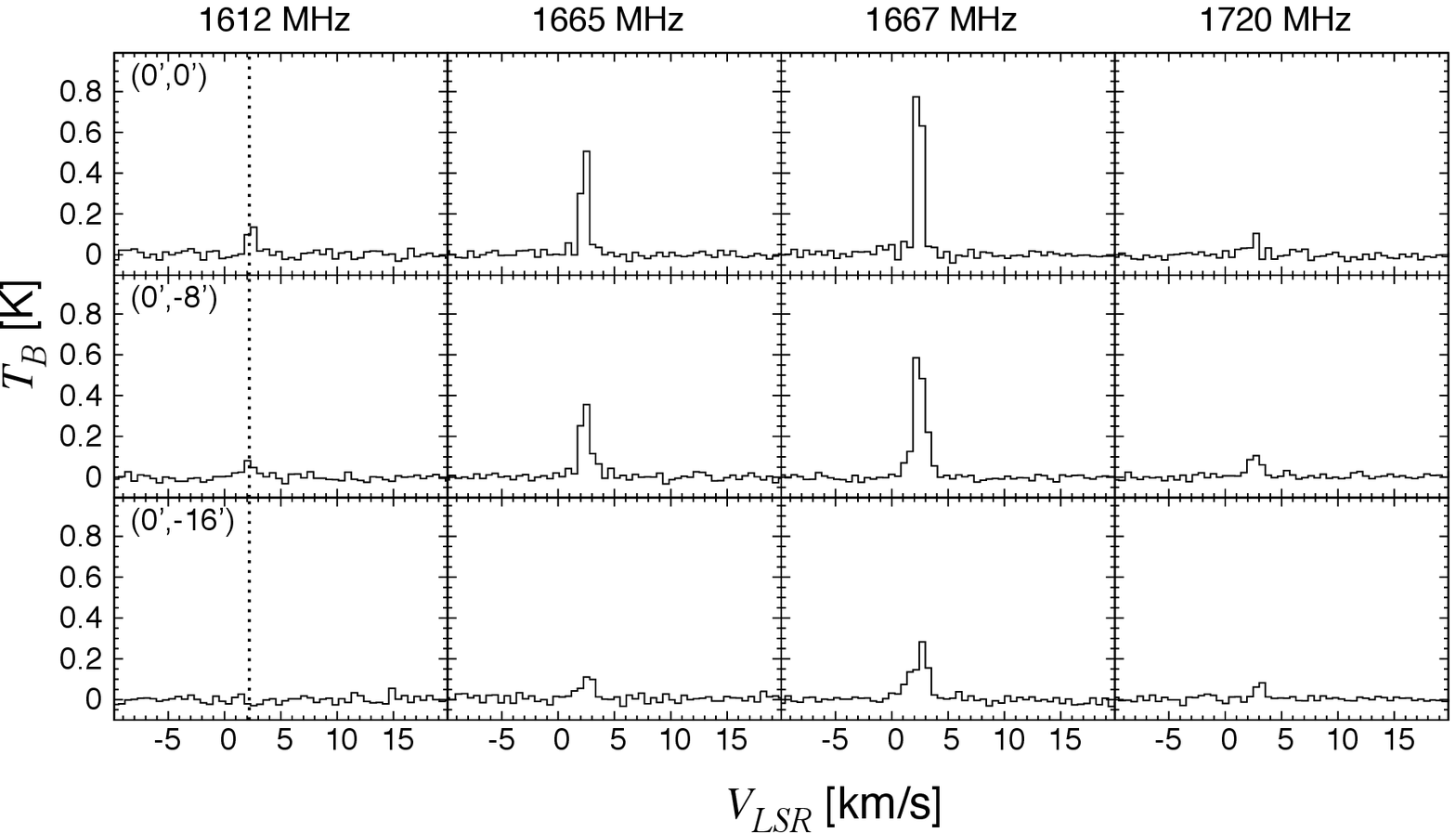}
  \caption{Observed spectra of the OH 18 cm transition toward L134N. A vertical dashed line represents the LSR velocity of the 1612 MHz line.
    \label{fig_spectrum_L183}}
\end{figure}

L134N is an isolated cold dark cloud without any associated stars.
A dense core traced by NH$_3$ exists in the northern part of the cloud and has been the subject of extensive astrochemical studies (e.g. Swade 1989; Dickens et al. 2000).
A rather diffuse molecular cloud traced by {}$^{13}$CO is extended toward the southern direction (Laureijs et al. 1995; Lehtinen et al. 2003).
We observed the OH 18 cm transition toward three positions with a spacing of $8'$, as shown in Figure $\ref{fig_map_L183}$.
Although the absorption feature of the 1612 MHz line is not clearly seen, the intensity of the 1612 MHz line becomes weaker toward the southern positions,
whereas the intensity of the 1720 MHz line relative to the main lines increases (Figure $\ref{fig_spectrum_L183}$).
The relative intensities of the four hyperfine components significantly deviate from the intrinsic line strengths,
{\it i.e.}, 1612:1665:1667:1720 MHz = 1:5:9:1 (Table $\ref{table_OH}$), and hence, they are clearly anomalous.
This trend can also be seen in the spectra observed toward the L134 cloud (1.5$^\circ$ south of L134N) by Mattila et al. (1979), although the signal-to-noise ratio of their spectra is rather poor.
At the southernmost position, a marginal absorption feature of the 1612 MHz line may be seen.

\subsection{$\rho$-Ophiuchi molecular cloud}
\begin{figure}
  \epsscale{1.1}
  \plotone{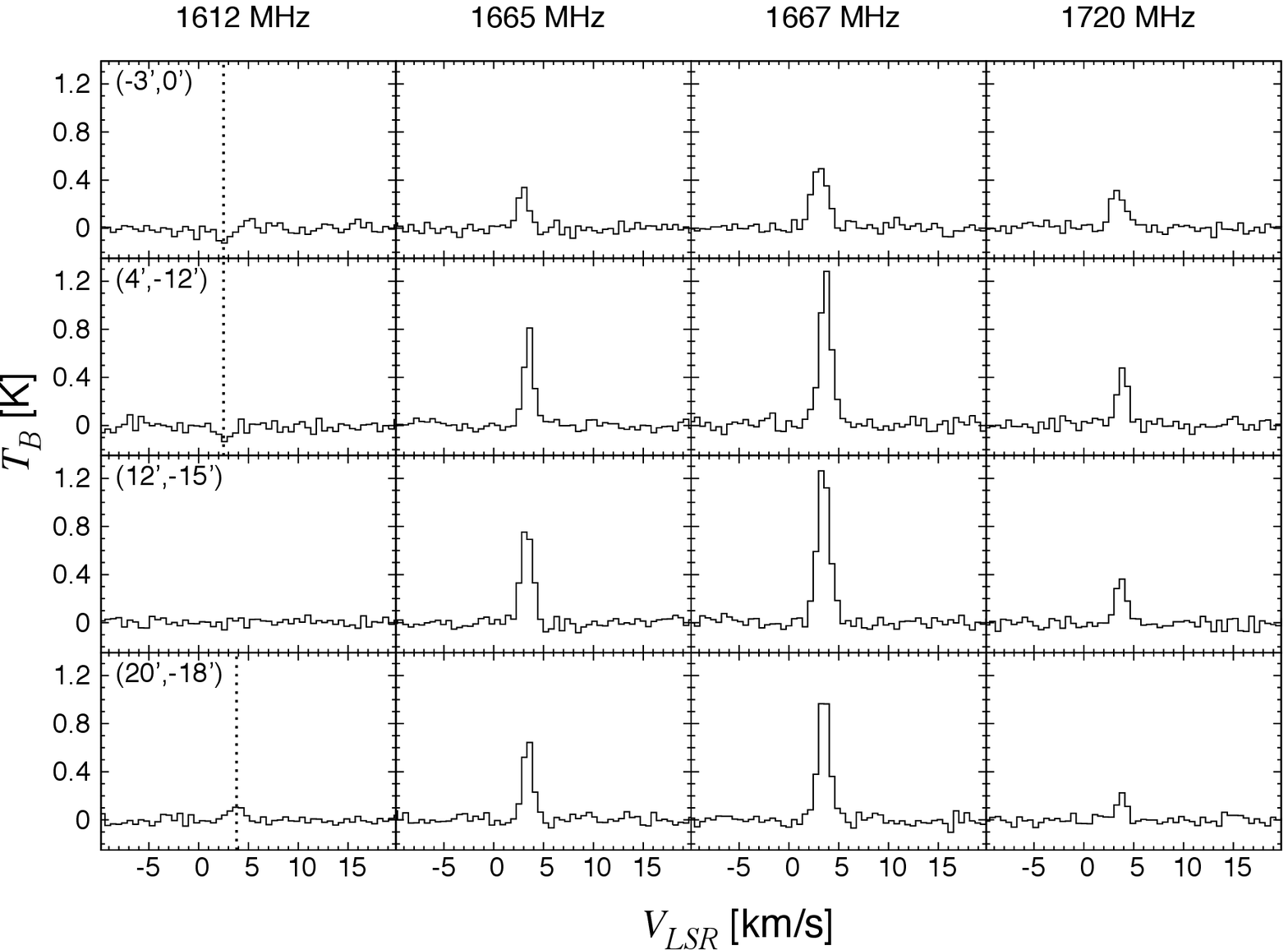}
  \caption{Observed spectra of the OH 18 cm transition toward $\rho$-Ophiuchi. A vertical dashed line represents the LSR velocity of the 1612 MHz line.
    \label{fig_spectrum_Oph}}
\end{figure}

The $\rho$-Ophiuchi molecular cloud is illuminated by the nearby B2 V star HD147889 (see Fig.$\ref{fig_map_Oph}$),
and a part of it constitutes a representative photodissociation region (PDR) (Yui et al. 1993; Liseau et al. 1999; Kamegai et al. 2003).
We observed four positions at different distances from HD147889.
Three of them are aligned on a straight line connecting HD147889 with the intensity peak of the [C I] emission (Kamegai et al. 2003),
while the other position is toward $\rho$-Oph A.
The 1612 MHz line appears in absorption toward the two positions nearest to HD147889 (Figure $\ref{fig_spectrum_Oph}$).
The absorption feature is less significant for the more detached positions, and it eventually changes to emission behind the [C I] peak position.
Line parameters of all spectra are summarized in Table $\ref{table_line_para}$.

\section{STATISTICAL EQUILIBRIUM CALCULATIONS} 
In order to constrain the physical conditions that are conducive for absorption to occur in the 1612 MHz component,
we conducted statistical equilibrium calculations assuming the photon-escape probability formalism for a static spherical cloud (Goldreich \& Kwan 1974, Osterbrock \& Ferland 2006).
We employed the collisional cross sections calculated by Offer et al. (1994), where the state-to-state collisional cross sections with ortho and para H$_2$ considering the fine and hyperfine structure levels are separately tabulated.
We ignored the overlapping effect of the hyperfine components, because the line width is as narrow as 0.5 km/s - 1.0 km/s for our target sources.
We confirmed that our program gives the same result as the statistical equilibrium calculation code RADEX (van der Tak et al. 2007).
\par
Figure $\ref{fig_lvg}$ (a) shows the intensities of the four hyperfine components of the ground rotational state as a function of the gas kinetic temperature at a H$_2$ density of 10$^4$ cm$^{-3}$, an OH column density of 5$\times$10$^{14}$ cm$^{-2}$, and an ortho-to-para (o/p) ratio of H$_2$ of 3.
At low temperatures below 20 K, the relative intensities of the hyperfine components are close to the LTE values.
However, the intensities of the 1612 MHz and 1720 MHz components become different above 20 K, where the conjugate behavior of these two components develops.
This trend is similar to that pointed out by Elitzur (1976).
As the temperature is further raised, the 1612 MHz line appears in absorption above 40 K.
Moreover, the main lines are also expected to be observed in absorption above 70 K.
Hence, the relative intensities of the hyperfine components are very sensitive to the gas kinetic temperature above 20 K.
The lower temperature limit for the anomaly originates from the condition that the collisional excitation to the first rotationally excited state is possible to some extent to cause non-LTE populations among hyperfine structure levels.
It should be noted that a deviation of the hyperfine intensities from the LTE values is also seen in the main line, which was pointed out by Crutcher (1979).
\par
On the other hand, the hyperfine intensity ratio is almost constant over a wide range of the H$_2$ density (10$^2$-–10$^7$ cm$^{-3}$) (Figure $\ref{fig_lvg}$ (b)).
Above 10$^7$ cm$^{-3}$, which is the critical density for excitation to the first rotationally excited state, intensities of the hyperfine components tend to approach to the LTE values.
The 1612 MHz absorption feature requires high OH column density, typically 10$^{14}$ cm$^{-2}$, for a gas kinetic temperature of 40 K or higher (Figure $\ref{fig_lvg}$ (c)).
In addition, it is worth noting that the intensities of the four hyperfine components are also sensitive to the o/p ratio of H$_2$, when the temperature is higher than 20 K.
If the o/p ratio is close to 0, not only the 1612 MHz line but also the main hyperfine components appear in absorption at 50 K (Figure $\ref{fig_lvg}$ (d)).
Hence, we can determine the gas kinetic temperature, the column density of OH, and the o/p ratio of H$_2$ from the intensities of the hyperfine components.

\section{RESULTS and DISCUSSIONS} 
We use our statistical equilibrium calculation code to derive the gas kinetic temperature and the column densities by least-square fitting the observed intensities of the four hyperfine components.
First, we analyze the spectrum averaged for the 8 positions observed in HCL2E (Figure $\ref{fig_average}$).
In this analysis, the H$_2$ density is fixed to 10$^3$ cm$^{-3}$, because it is insensitive to the other parameters over quite a wide range from 10$^2$ to 10$^7$ cm$^{-3}$.
The least-squares fit on the four hfs components yields an OH column density of (4.4 $\pm$ 0.3) $\times$ 10$^{14}$ cm$^{-2}$,
a gas kinetic temperature of 60 $\pm$ 3 K, a value of 3.5 $\pm$ 0.9 for an o/p ratio of H$_2$.
This result does not change within the error, if the H$_2$ density varies between 10$^2$ cm$^{-3}$ or 10$^4$ cm$^{-3}$.
The obtained parameters well reproduce the observed spectra (Figure $\ref{fig_average}$).
It should be noted that the quoted errors do not include the systematic uncertainty due to the collisional rates used, the radiative transfer model,
and the effect of neglecting infrared excitations to the excited rotational states. (e.g. Litvak et al. 1969)
\par
\begin{figure}
  \epsscale{0.7}
  \plotone{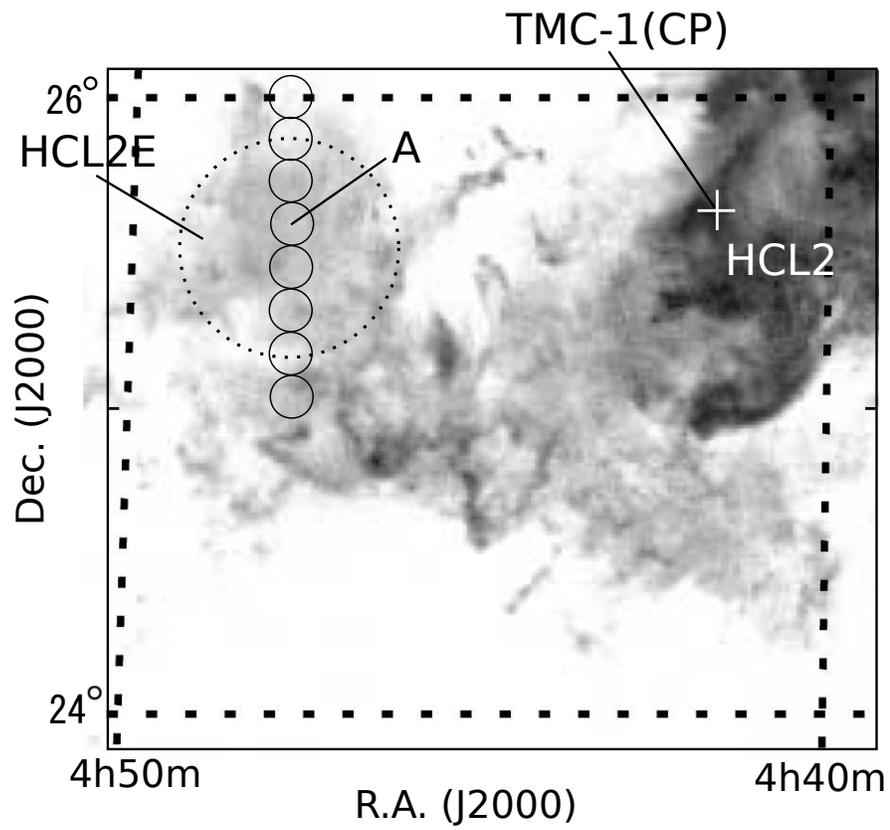}
  \caption{The peak intensity image of the {}$^{13}$CO ($J$=1-0) line (Narayanan et al. 2008). Circles represent the observed positions in HCL2E.
    `A' is the `central' position, which is a peak of the [C I] intensity (Maezawa 2000).
    A cross mark represents the position of TMC-1 cyanopolyyne peak (CP).
    \label{fig_map_HCL2E}}
\end{figure}
\begin{figure}
  \epsscale{1}
  \plotone{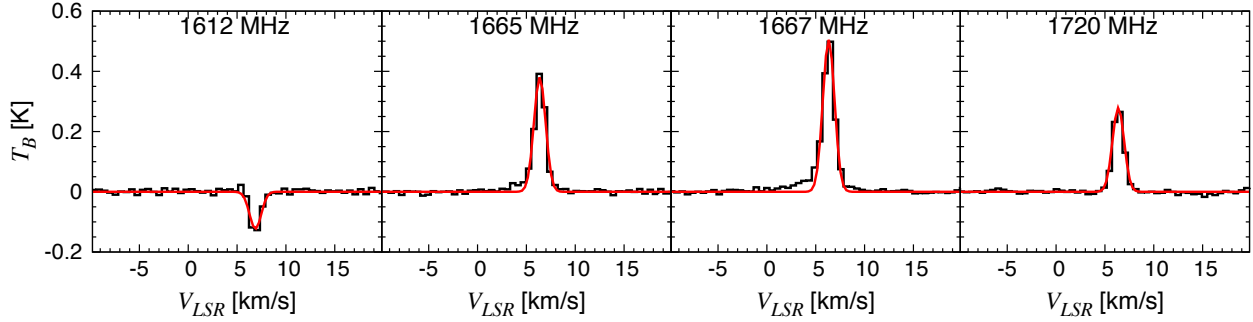}
  \caption{Spectra for the four OH hyperfine structure lines produced by averaging the eight spectra (Fig. $\ref{fig_map_HCL2E}$) each observed toward the positions marked in Fig. $\ref{fig_map_HCL2E}$.
  Red lines show the Gaussian profiles with the best fit parameters, namely an OH column density of $4.4 \times 10^{14}$ cm$^{-2}$, a gas kinetic temperature of 60 K, and an ortho-to-para ratio of 3.5.\label{fig_average}}
\end{figure}
\begin{figure}
  \plottwo{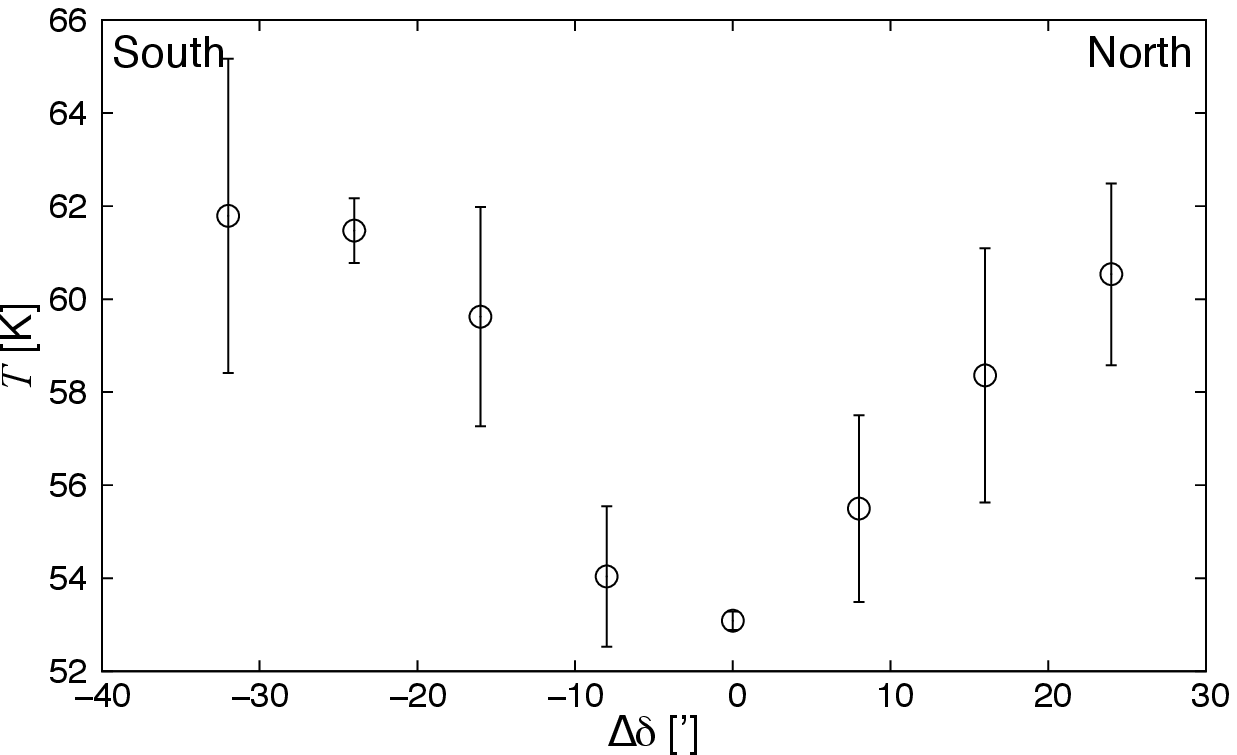}{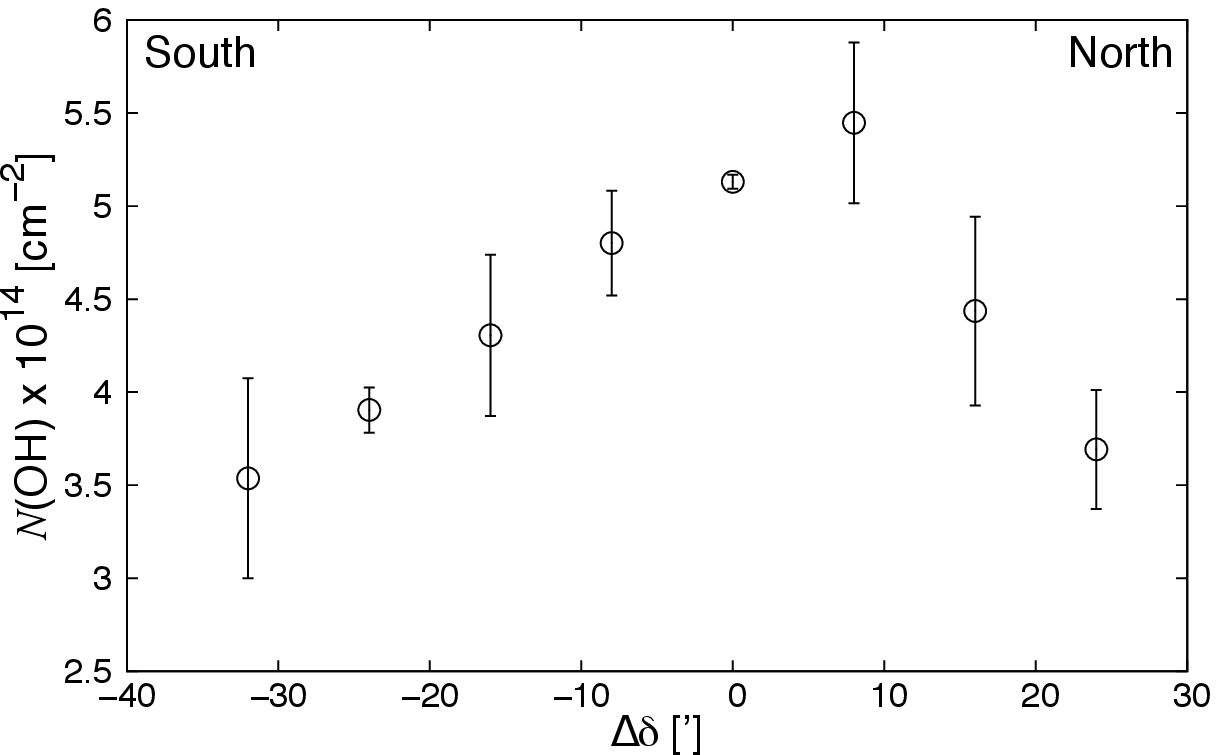}
  \caption{Gas kinetic temperatures (left) and column densities of OH (right) for the eight positions observed in HCL2E derived from the statistical equilibrium calculations
    as a function of declination angluar offset.
    Error bars show three times the standard deviation. \label{fig_pos_vs_T_OH}}
\end{figure}

Further, we analyze individual positions to derive the OH column density and the gas kinetic temperature.
The o/p ratio derived above is close to the statistical value of 3, and hence, we fix it to 3 in the fit.
The results are shown in Table $\ref{table_phys_para}$.
For HCL2E, the derived gas kinetic temperature is 53 $\pm$ 1 K toward the central position (position A in Figure $\ref{fig_map_HCL2E}$), and rises to 62 $\pm$ 3 K for positions located closer to the cloud boundary (Figure $\ref{fig_pos_vs_T_OH}$).
Since HCL2E is a translucent cloud whose visual extinction is about 4 magnitude (Maezawa 2000), this is naturally interpreted as photoelectric heating of the gas under irradiation of the interstellar UV radiation.
In contrast, the column density of OH is as high as (4--5) $\times$ 10$^{14}$ cm$^{-2}$ toward the central three positions, and decreases to 3 $\times$ 10$^{14}$ cm$^{-2}$ at the peripheries (Figure $\ref{fig_pos_vs_T_OH}$).
Toward the central position, the fractional abundance of OH is estimated to be 1 $\times$ 10$^{-7}$.
This is a typical fractional abundance of OH reported for diffuse and translucent clouds (Wiesemeyer et al. 2012; Weselak et al. 2010; Felenbok \& Roueff 1996),
and is consistent with chemical models (e.g. Le Petit et al. 2004).
\par
\begin{center}
\begin{figure}
  \begin{center}
  \epsscale{1.1}
  \plotone{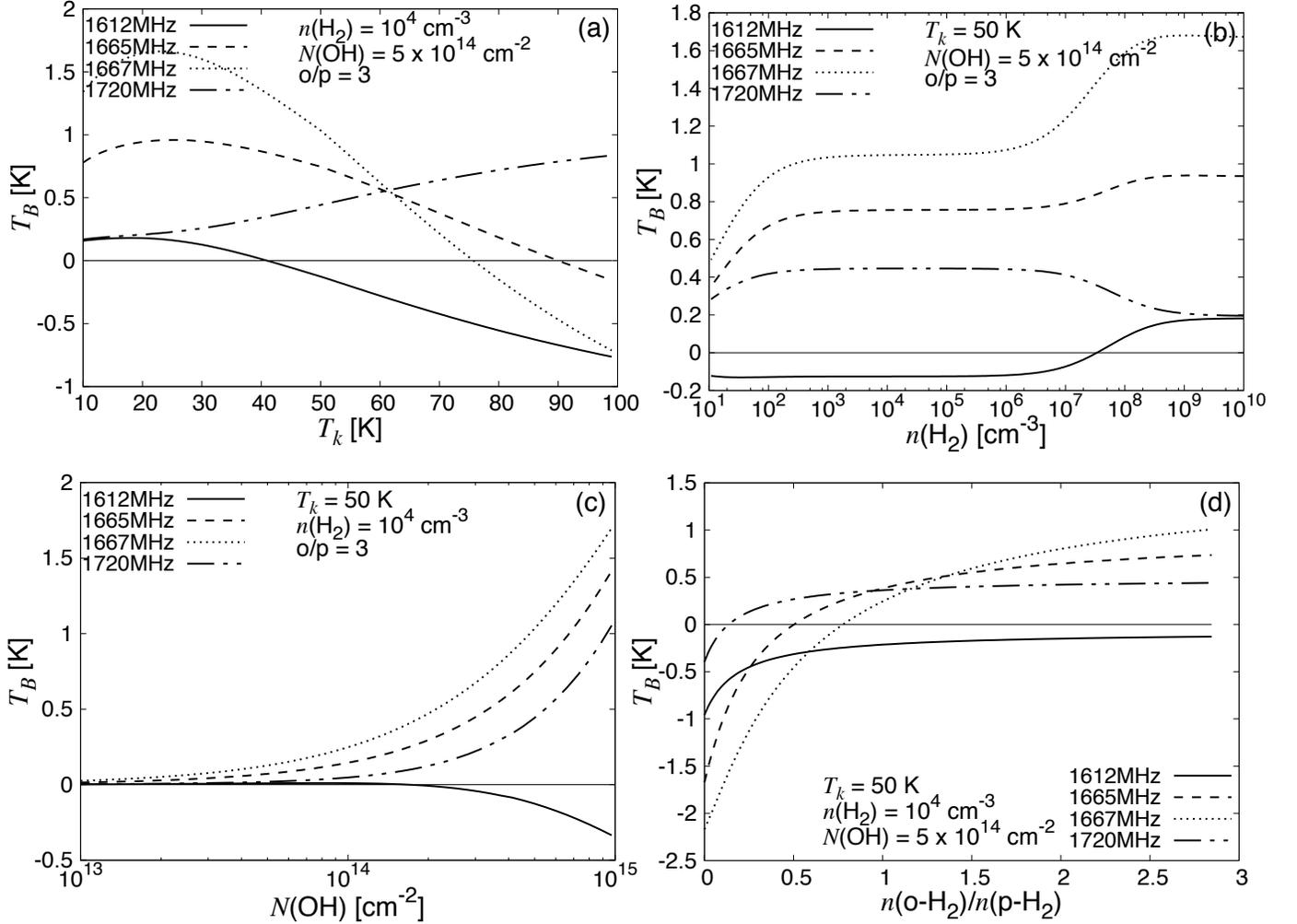}
  \caption{The expected intensities of the OH hyperfine structure lines derived from the statistical equilibrium calculations
    as a function of the gas kinetic temperature ($T_k$) (a), H$_2$ density ($n({\rm H_2})$) (b),
    the column density of OH ($N$(OH)) (c) and the ortho-to-para (o/p) ratio of H$_2$ (d).\label{fig_lvg}
  }
\end{center}
\end{figure}
\end{center}

A similar least-squares analysis is also carried out toward the dark cloud L134N.
Since the absorption feature is not clearly seen in this source, the gas kinetic temperature of the source traced by the OH line must be lower than 40 K.
Toward the dense core position, the intensity anomaly is almost absent.
This means that the gas kinetic temperature is low and the excitation to the first rotationally excited state is inefficient.
According to Figure $\ref{fig_lvg}$, the gas kinetic temperature must be lower than 30 K for the intensities of the 1612 MHz line and the 1720 MHz line to be identical.
We could reproduce the observed hyperfine pattern with a gas kinetic temperature below 25 K.
Table $\ref{table_phys_para}$ lists the result for a gas kinetic temperature of 10 K, which is the typical temperature for starless cores (e.g. Benson \& Myers 1989).
It should be noted that the 1720 MHz line may look slightly brighter than the 1612 MHz line.
This is similar to what is reported for TMC-1 by Harju et al. (2000).
However, it cannot be explained by our statistical equilibrium calculations, and more sophisticated analyses appear to be required.
For the other two positions of L134N, the gas kinetic temperature is determined from the hyperfine intensities.
The temperature tends to increase toward the periphery, as shown in Table $\ref{table_phys_para}$, indicating a contribution of the photoelectric heating by the interstellar UV radiation.
\par
\begin{figure}
  \epsscale{0.6}
  \plotone{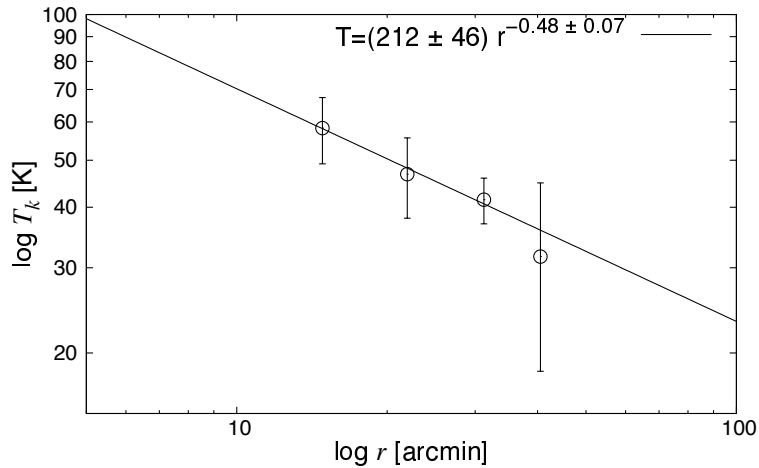}
  \caption{Gas kinetic temperatures for the four positions observed in $\rho$-Ophiuchi derived from the statistical equilibrium calculations as a function of the apparent distance ($r$) from the
  exciting star HD147889 (Figure $\ref{fig_map_Oph}$).
  Error bars show three times the standard deviation.
  The temperature approximately follows the $r^{-1/2}$ law.\label{fig_dis_vs_T_OH}}
\end{figure}

Since the intensity anomaly of the hyperfine components is clearly seen toward the four positions of the $\rho$-Ophiuchi molecular cloud, the gas kinetic temperature of each position was determined by least-squares analysis.
This region is a photodissociation region illuminated by the exciting star HD147889.
Weak radio continuum emission is extended over this region.
However, its brightness temperature is mostly less than 100 mK at 2.3 GHz (Yui et al. 1993; Baart et al. 1980) which is much lower than the cosmic microwave background temperature, and hence, we ignore it in the analysis.
Furthermore, we assume an o/p ratio of H$_2$ to be the statistical value of 3.
The gas kinetic temperature and the column density derived for the four positions are summarized in Table $\ref{table_phys_para}$.
The temperature is found to decrease with increasing distance, $r$, from the exciting star with the $r^{-1/2}$ law (Figure $\ref{fig_dis_vs_T_OH}$), which is expected from the balance between the heating by the exciting star and thermal cooling.
Kamegai et al. (2003) measured the excitation temperature of the fine structure levels of the neutral carbon atom to be 38 K and 29 K toward Peak I and the [C I] peak (Peak II) positions, respectively, by observing the ${}^3P_1-{}^3P_0$ (492 GHz) and ${}^3P_2-{}^3P_1$ (809 GHz) lines.
These temperatures are lower than the present estimate from the OH hyperfine anomaly, probably because the [C I] lines trace denser regions in the PDR.
\par
As demonstrated in this paper, the hyperfine intensity pattern of the OH 18 cm transition is an important probe for characterizing physical conditions of peripheries/envelopes of molecular clouds.
In particular, it is a good thermometer over a wide range of the H$_2$ density.
It is a good tracer of the warm envelope of molecular cloud which has not been studied extensively with the lines of CO and its isotopic species or the [H I] 21 cm line.
These novel aspects of the OH 18 cm transition will be useful to trace transition zones between molecular cloud cores and diffuse clouds,
which is important for an understanding of the formation processes of molecular clouds.
However, the behavior of the hyperfine components of the OH 18 cm transition has not fully been understood.
For example, a lower intensity of the 1720 MHz line than the 1612 MHz line cannot be explained by the statistical equilibrium calculations presented here.
Such an anomaly is also seen in TMC-1 (Harju et al. 2000) and W40 (Crutcher et al. 1977).
We shall present our more extensive observational exploration of this anomaly, together with further modeling, in a separate publication.

\acknowledgements
We thank the staff of the MPIfR 100 m telescope at Effelsberg for excellent support.
This study is supported by Grants-in-Aid from Ministry of Education, Sports, Science, and Technologies of Japan
(21224002, 25400223, and 25108005).

\appendix
\clearpage
\begin{deluxetable}{ccc}
  \tablewidth{0pt}
  \tablecaption{Frequencies and line strength of the OH 18 cm transition (${}^2{\Pi}_{3/2}, J=3/2$) \label{table_OH}}
  \tablehead{%
    Transition&$\nu$ [MHz]\tablenotemark{a}&S${\mu}^2$\ \tablenotemark{b}
  }
  \startdata
  $F=1-2$&1612.2310&0.79\\
  $F=1-1$&1665.4018&4.0\\
  $F=2-2$&1667.3590&7.1\\
  $F=2-1$&1720.5300&0.79
  \enddata
  \tablenotetext{a}{Taken from Darling (2003)}
  \tablenotetext{b}{Calculated from the Einstein coefficients tabulated in Leiden Atomic and Molecular Database (Sch\"oier et al. 2005)}
\end{deluxetable}
\begin{deluxetable}{cccccccccccccc}
  \tabletypesize{\tiny}
  \rotate
  \tablewidth{0pt}
  \tablecaption{Observed line parameters\label{table_line_para}}
  \tablehead{%
    &1612 MHz&&&1665MHz&&&1667 MHz&&&1720MHz&&\\
    Position&${T_B}[K]$&$\Delta V$[km/s]&$V_{LSR}[km/s]$&${T_B}[K]$&$\Delta V$[km/s]&$V_{LSR}[km/s]$&${T_B}[K]$&$\Delta V$[km/s]&$V_{LSR}[km/s]$&${T_B}[K]$&$\Delta V$[km/s]&$V_{LSR}[km/s]$
  }
\startdata
HCL2E&&&&&&&&&&&&\\
(0$'$,+24$'$)& -0.10(1)& 1.37(4)& 6.77(10)& 0.31(1)& 1.37(4)& 6.30(3)& 0.37(1)& 1.37(4)& 6.14(2)& 0.22(1)& 1.37(4)& 6.22(4)\\
(0$'$,+16$'$)& -0.15(1)& 1.38(4)& 6.97(8)& 0.40(1)& 1.38(4)& 6.30(2)& 0.49(1)& 1.38(4)& 6.21(2)& 0.26(1)& 1.38(4)& 6.36(4)\\
(0$'$,+8$'$)& -0.14(1)& 1.51(3)& 6.86(8)& 0.48(1)& 1.51(3)& 6.16(2)& 0.62(1)& 1.51(3)& 6.12(2)& 0.30(1)& 1.51(3)& 6.22(4)\\
(0$'$,0$'$)& -0.13(2)& 1.24(3)& 6.84(11)& 0.57(2)& 1.24(3)& 6.24(2)& 0.78(2)& 1.24(3)& 6.22(2)& 0.36(2)& 1.24(3)& 6.16(4)\\
(0$'$,-8$'$)& -0.17(1)& 1.06(2)& 6.70(6)& 0.59(2)& 1.06(2)& 6.41(2)& 0.82(2)& 1.06(2)& 6.38(1)& 0.42(2)& 1.06(2)& 6.29(2)\\
(0$'$,-16$'$)& -0.17(1)& 1.26(4)& 6.87(7)& 0.37(2)& 1.26(4)& 6.57(3)& 0.50(2)& 1.26(4)& 6.55(2)& 0.31(1)& 1.26(4)& 6.52(3)\\
(0$'$,-24$'$)& -0.11(2)& 1.45(6)& 6.99(13)& 0.28(2)& 1.45(6)& 6.64(5)& 0.36(2)& 1.45(6)& 6.64(4)& 0.22(2)& 1.45(6)& 6.65(6)\\
(0$'$,-32$'$)& -0.13(2)& 1.28(6)& 6.93(10)& 0.26(2)& 1.28(6)& 6.64(5)& 0.38(2)& 1.28(6)& 6.61(4)& 0.24(2)& 1.28(6)& 6.57(6)\\
\tableline
L134N&&&&&&&&&&&&\\
(0$'$,0$'$)& 0.18(2)& 0.74(3)& 2.38(4)& 0.60(3)& 0.74(3)& 2.35(1)& 1.03(4)& 0.74(3)& 2.40(1)& 0.11(2)& 0.74(3)& 2.62(8)\\
(0$'$,-8$'$)& 0.08(1)& 1.21(3)& 2.15(13)& 0.35(1)& 1.21(3)& 2.40(3)& 0.62(2)& 1.21(3)& 2.43(2)& 0.12(1)& 1.21(3)& 2.63(8)\\
(0$'$,-16$'$)& -0.02(1)& 1.67(10)& 2.81(46)& 0.11(1)& 1.67(10)& 2.57(11)& 0.25(1)& 1.67(10)& 2.55(5)& 0.06(1)& 1.67(10)& 3.09(18)\\
\tableline
$\rho$-Oph&&&&&&&&&&&&\\
(-3$'$,0$'$)& -0.12(2)& 1.70(8)& 2.51(20)& 0.29(2)& 1.70(8)& 2.94(8)& 0.54(3)& 1.70(8)& 3.08(4)& 0.32(2)& 1.70(8)& 3.39(7)\\
(+4$'$,-12$'$)& -0.14(3)& 1.31(3)& 2.68(16)& 0.74(3)& 1.31(3)& 3.52(3)& 1.29(3)& 1.31(3)& 3.68(2)& 0.47(3)& 1.31(3)& 3.87(5)\\
(+12$'$,-15$'$)& -0.02(2)& 1.44(3)& 2.12(88)& 0.79(2)& 1.44(3)& 3.32(3)& 1.32(3)& 1.44(3)& 3.46(2)& 0.35(2)& 1.44(3)& 3.72(6)\\
(+20$'$,-18$'$)& 0.13(2)& 1.29(3)& 3.82(14)& 0.65(2)& 1.29(3)& 3.43(3)& 1.10(3)& 1.29(3)& 3.51(2)& 0.21(2)& 1.29(3)& 3.77(8)\\
  \enddata
  \tablecomments{$T_B$, $\Delta V$ and $V_{LSR}$ are obtained by a Gaussian fit, assuming that the $\Delta V$ values of the four hyperfine components are the same for the same observed position. The error in the parentheses represents one standard deviation.}
\end{deluxetable}
\begin{deluxetable}{cccccc}
  \tabletypesize{\small}
  \tablewidth{0pt}
  \tablecaption{Derived parameters (3 sources)\label{table_phys_para}}
  \tablehead{%
    source&
    position&
    $T_k$ [K]&
    $N$(OH) [cm$^{-2}$]&
    o/p ratio
  }
  \startdata
  HCL2E&(0$'$,24$'$)\tablenotemark{a}&61(2)&3.7(3)$\times$10$^{14}$&3(fixed)\\
  HCL2E&(0$'$,16$'$)&58(3)&4.4(5)$\times$10$^{14}$&3(fixed)\\
  HCL2E&(0$'$,8$'$)&56(2)&5.4(4)$\times$10$^{14}$&3(fixed)\\
  HCL2E&(0$'$,0$'$)&53(1)&5.1(1)$\times$10$^{14}$&3(fixed)\\
  HCL2E&(0$'$,-8$'$)&54(2)&4.8(3)$\times$10$^{14}$&3(fixed)\\
  HCL2E&(0$'$,-16$'$)&60(2)&4.3(4)$\times$10$^{14}$&3(fixed)\\
  HCL2E&(0$'$,-24$'$)&61(1)&3.9(1)$\times$10$^{14}$&3(fixed)\\
  HCL2E&(0$'$,-32$'$)&62(3)&3.5(5)$\times$10$^{14}$&3(fixed)\\
  L134N&(0$'$,0$'$)&10(fixed)\tablenotemark{b}&3.0(5)$\times$10$^{14}$&0(fixed)\\
  L134N&(0$'$,-8$'$)&33(24)&2.3(6)$\times$10$^{14}$&3(fixed)\\
  L134N&(0$'$,-16$'$)&57(16)&2.0(13)$\times$10$^{14}$&3(fixed)\\
  $\rho$-Oph&(-3$'$,0$'$)&58(9)&5.3(20)$\times$10$^{14}$&3(fixed)\\
  $\rho$-Oph&(+4$'$,-12$'$)&46(9)&7.1(18)$\times$10$^{14}$&3(fixed)\\
  $\rho$-Oph&(+12$'$,-15$'$)&41(5)&7.1(8)$\times$10$^{14}$&3(fixed)\\
  $\rho$-Oph&(+20$'$,-18$'$)&32(13)&4.4(7)$\times$10$^{14}$&3(fixed)\\

  \enddata
  \tablenotetext{a}{The ($0'$,$0'$) positions are as follows: %
    For HCL2E, ($\alpha_{2000}, \delta_{2000}$) = ($4^h 48^m 15^s.1$, $25^{\circ} 35' 15''$), which corresponds to position A of Figure $\ref{fig_map_HCL2E}$; %
  for L134N, ($\alpha_{2000}, \delta_{2000}$) = ($15^h 54^m 0^s.5$, $ -2^{\circ} 51' 49''$); %
  for $\rho$-Ophiuchi, ($\alpha_{2000}, \delta_{2000}$) = ($16^h 26^m 31^s.4$, $-24^{\circ} 21' 44''$).%
}
\tablenotetext{b}{If $T_k$ is assumed to be 20 K, $N$(OH) is (3.1 $\pm$ 0.6) $\times$ 10$^{14}$ cm$^{-2}$}
\end{deluxetable}
\end{document}